\documentclass{emulateapj}
\usepackage{url}

\shorttitle{Progenitor of long-duration GRB 211227A}
\shortauthors{L\"{u} et al.}
\slugcomment{}

\begin{document}

\title{GRB 211227A as a peculiar long gamma-ray burst from compact star merger}
\author{Hou-Jun L\"{u}\altaffilmark{1}, Hao-Yu Yuan\altaffilmark{1}, Ting-Feng Yi\altaffilmark{2},
Xiang-Gao Wang\altaffilmark{1}, You-Dong Hu\altaffilmark{3}, Yong Yuan\altaffilmark{4}, Jared
Rice\altaffilmark{5}, Jian-Guo Wang\altaffilmark{6}, Jia-Xin Cao\altaffilmark{1}, De-Feng
Kong\altaffilmark{1}, Emilio Fernandez-Garc\'ia\altaffilmark{3}, Alberto
J.Castro-Tirado\altaffilmark{3,7}, Ji-Shun Lian\altaffilmark{1}, Wen-Pei Gan\altaffilmark{1},
Shan-Qin Wang\altaffilmark{1}, Li-Ping Xin\altaffilmark{8}, M.~D.
Caballero-Garc\'{i}a\altaffilmark{3}, Yu-Feng Fan\altaffilmark{6}, and En-Wei
Liang\altaffilmark{1}} \altaffiltext{1}{Guangxi Key Laboratory for Relativistic Astrophysics,
School of Physical Science and Technology, Guangxi University, Nanning 530004, China;
lhj@gxu.edu.edu; lew@gxu.edu.cn} \altaffiltext{2}{Key Laboratory of Colleges and Universities in
Yunnan Province for High-energy Astrophysics, Department of Physics, Yunnan Normal University,
Kunming 650500, China} \altaffiltext{3}{Instituto de Astrof\'isica de Andaluc\'ia (IAA-CSIC),
Glorieta de la Astronom\'ia s/n, E-18008, Granada, Spain}\altaffiltext{4}{School of Physics Science
And Technology ,Wuhan University No.299 Bayi Road, Wuhan, Hubei, China}\altaffiltext{5}{Department
of Physics, Texas State University, San Marcos, TX 78666, USA}\altaffiltext{6}{Yunnan
Observatories, Chinese Academy of Sciences, Kunming, China}\altaffiltext{7}{Unidad Asociada al CSIC
Departamento de Ingenier\'ia de Sistemas y Autom\'atica, Escuela de 16 Ingenier\'ia Industrial,
Universidad de M\'alaga, M\'alaga, Spain}\altaffiltext{8}{CAS Key Laboratory of Space Astronomy and
Technology, National Astronomical Observatories, Chinese Academy of Sciences, Beijing 100101,
China}

\begin{abstract}	
Long-duration gamma-ray bursts (GRBs) associated with supernovae (SNe) are believed to originate
from massive star core-collapse events, whereas short-duration GRBs that are related to compact
star mergers are expected to be accompanied by kilonovae. GRB 211227A, which lasted about 84 s, had
an initial short/hard spike followed by a series of soft gamma-ray extended emission at redshift
$z=$0.228. We performed follow-up observations of the optical emission using BOOTES, LCOGT, and the
Lijiang 2.4m telescope, but we detected no associated supernova signature, even down to very
stringent limits at such a low redshift. We observed the host galaxy within a large error circle
and roughly estimate the physical offset of GRB 211227A as $20.47\pm14.47$ kpc from the galaxy
center. These properties are similar to those of GRB 060614, and suggest that the progenitor of GRB
211227A is not favored to be associated with the death of massive stars. Hence, we propose that GRB
211227A originates from a compact star merger. Calculating pseudo-kilonova emission for this case
by adopting the typical parameters, we find that any associated pseudo-kilonova is too faint to be
detected. If this is the case, it explains naturally the characteristics of the prompt emission,
the lack of SN and kilonova emission, and the large physical offset from the galaxy center.
\end{abstract}


\section{Introduction}
Gamma-ray bursts (GRBs) are thought to originate in violent events, such as massive star
core-collapse or the compact star mergers (Woosley 1993; Pacz\'{y}nski 1986; also see Kumar \&
Zhang 2015 for a review). Such catastrophic destruction of these progenitor systems may result in
the formation of a magnetar or black hole which powers a relativistic jet pointing in the direction
of the observer (e.g., Eichler et al. 1989; Usov 1992; Thompson 1994; Dai \& Lu 1998a,b; Popham et
al. 1999; Narayan, et al. 2001; Zhang \& M{\'e}sz{\'a}ros 2001; Lei et al. 2009; Metzger et al.
2011; Bucciantini et al. 2012; L\"{u} \& Zhang 2014; Berger 2014). Within the standard fireball
model scenario, the observed variability of $\gamma$-ray emission is caused by photosphere emission
(Thompson 1994; Ghisellini \& Celotti 1999; Pe'er et al. 2006; Beloborodov 2010; Lazzati \&
Begelman 2010), internal shocks (Rees \& M{\'e}sz{\'a}ros 1994; Sari \& Piran 1997), or Internal
Collision-induced Magnetic Reconnection and Turbulence (ICMART; Zhang \& Yan 2011). A
multi-wavelength afterglow emission is attributed to the external shock when the fireball is
decelerated by a sufficient amount of external material (M{\'e}sz{\'a}ros \& Rees 1997; Sari et al.
1998).

Phenomenally, GRBs are divided as long- and short-duration with a division line at the observed
duration $T_{90}\sim$2 s (Kouveliotou et al. 1993). From the theoretical point of view,
long-duration GRBs are powered by a relativistic jet that breaks out of the envelope of a massive
star when it undergoes core-collapse. There is about $\rm 0.1 M_{\odot}$ of nickel that can also be
created via explosive nucleosynthesis during this collapse. Therefore, a bright optical/infrared
transient called a supernova (SN; Woosley \& Bloom 2006; Hjorth \& Bloom 2012) should be produced
when the $\rm^{56}Ni$ decays into cobalt. Naturally, some long-duration GRBs are associated with SN
when the distance is not large enough (Galama et al. 1998; Kippen et al. 1998; Hjorth et al. 2003;
Stanek et al. 2003; Malesani et al. 2004; Della Valle et al. 2006; Mazzali et al. 2008; Bufano et
al. 2012; Xu et al. 2013; Cano et al. 2017; L\"{u} et al. 2018). On the other hand, the
short-duration GRBs are believed to originated from compact star mergers. The leading candidates
are neutron star$-$neutron star (NS$-$NS) and neutron star$-$black hole (NS$-$BH) systems. The
NS$-$NS merger may result in a magnetar as remnant (Rosswog et al. 2000; Dai et al. 2006; Metzger
et al. 2010; Yu et al. 2013; Zhang 2013; Lasky et al. 2014; L\"{u} et al. 2015) or black hole
(Rosswog et al. 2014). If this is the case, a mildly isotropic, sub-relativistic ejecta with
$10^{-4}-10^{-2} M_{\odot}$ and 0.1$-$0.3$c$, can be ejected during the merger to heat ejecta
(Rosswog et al. 2014; Hotokezaka et al. 2013). So that, an optical/infrared transient that is
powered by radioactive decay from r-process radioactive materials may be detected from any
direction if the flux is high enough (Li \& Pacz\'{y}nski 1998; Kulkarni 2005; Metzger et al. 2010;
Barnes \& Kasen 2013), and the transient is known as a macronova, kilonova, or merger-nova
(Kulkarni 2005; Metzger et al. 2010; Yu et al. 2013).

Observationally, the first GRB-SN association event was discovered as the under$-$luminous GRB
980425 and the Type Ic SN 1998bw at redshift $z=$0.0085 (Galama et al. 1998; Kippen et al. 1998;
Pian et al. 1998; Sadler et al. 1998). Afterwards, a handful of long-duration GRBs associated with
spectroscopically$-$identified type Ib/c SNe were detected (Kovacevic et al. 2014; Cano et al.
2016; L\"u et al. 2018). By comparing with the first GRB$-$SN association, the first kilonova
emission event from a compact star merger was discovered in GRB 130603B with an excess near$-$IR
emission matching the predictions for r-process emission (Tanvir et al. 2013; Berger et al. 2013;
Fan et al. 2013; {Fong et al. 2014}). After that, several short GRBs were claimed to be associated
with a kilonova or mergernova, e.g., GRB 050709 (Jin et al. 2016), GRB 060614 (Yang et al. 2015),
GRB 070809 (Jin et al. 2020), GRB 080503 (Gao et al. 2015), GRB 160821B (Kasliwal et al. 2017;
Troja et al. 2019; Lamb et al. 2019), and GRB 150101B (Troja et al. 2018). Gao et al (2017) carried
out a complete search for magnetar$-$powered mergernovae from a sample of {\em Swift} GRBs, and
found that three magnetar$-$powered mergernova candidates are associated with short GRBs (050724,
061006 and 070714B) from late optical observations. Rastinejad et al. (2021) presented a
comprehensive optical and near-infrared catalog to search for possible kilonova emission and
constrain its ejecta mass (also see Yuan et al. 2021).

A particularly interesting case, GRB 060614 which is a nearby long-duration GRB at $z=0.125$
(Gehrels et al. 2006; Gal-Yam et al. 2006; Fynbo et al. 2006; Della Valle et al. 2006), needs to be
mentioned again. The light curve of GRB 060614 is characterized by a short/hard spike (with a
duration $\sim 5$ s) followed by a series of soft gamma-ray extended emission\footnote{The observed
extended emission is softer than that of initial short/hard spike of short GRBs and gamma-ray
emission of typical long GRBs. The extended emission detected in the BAT band seems to be simply
the internal plateau emission when the emission is bright and hard enough (L\"{u} et al. 2015).}
with a duration $\sim 100$ s. Phenomenologically, it definitely belongs to the long-duration GRB
population and a bright SN was expected. It should have been detected at such a low redshift, but a
SN association was not detected even with a deep search. These facts are consistent with the
compact star merger scenario, but they are not the direct evidence (Gehrels et al. 2006; Zhang et
al. 2007a). There was an ongoing debate on the physical origin of this case until 2015. Possible
evidence of a kilonova from SN-less long-duration GRB 060614 was claimed (Yang et al. 2015). They
discovered a near-infrared bump (only two data points) that is significantly above the regular
decaying afterglow, and therefore claimed it to be a signature of the kilonova component.

GRB 211227A is potentially associated with a galaxy at redshift $z=$0.228 (Beardmore et al. 2021;
Malesani et al. 2021). The light curve of its prompt emission is very similar to that of GRB
060614, which also has no SN association at this low redshift. However, the difference with GRB
060614 is that we do not find any kilonova signature with GRB 211227A. One question is what is the
progenitor of GRB 211227A, massive star collapse or compact star merger? In this paper, we
systematically analyze the observational data of both prompt emission, afterglow, and host galaxy
(in \S 2). Then, comparisons to GRB 211227A and GRB 060614 are shown in \S 3. In \S 4, we attempt
to investigate the possible origin by calculating the SN and kilonova emissions. The conclusions
are drawn in \S 5 with some additional discussion. Throughout the paper, a concordance cosmology
with parameters $H_0=71~\rm km~s^{-1}~Mpc^{-1}$, $\Omega_M=0.30$, and $\Omega_{\Lambda}=0.70$ is
adopted.

\section{The observations and data analysis}
\subsection{{\em Swift} BAT observations}
\begin{figure}
\centering
\includegraphics [angle=0,scale=0.6] {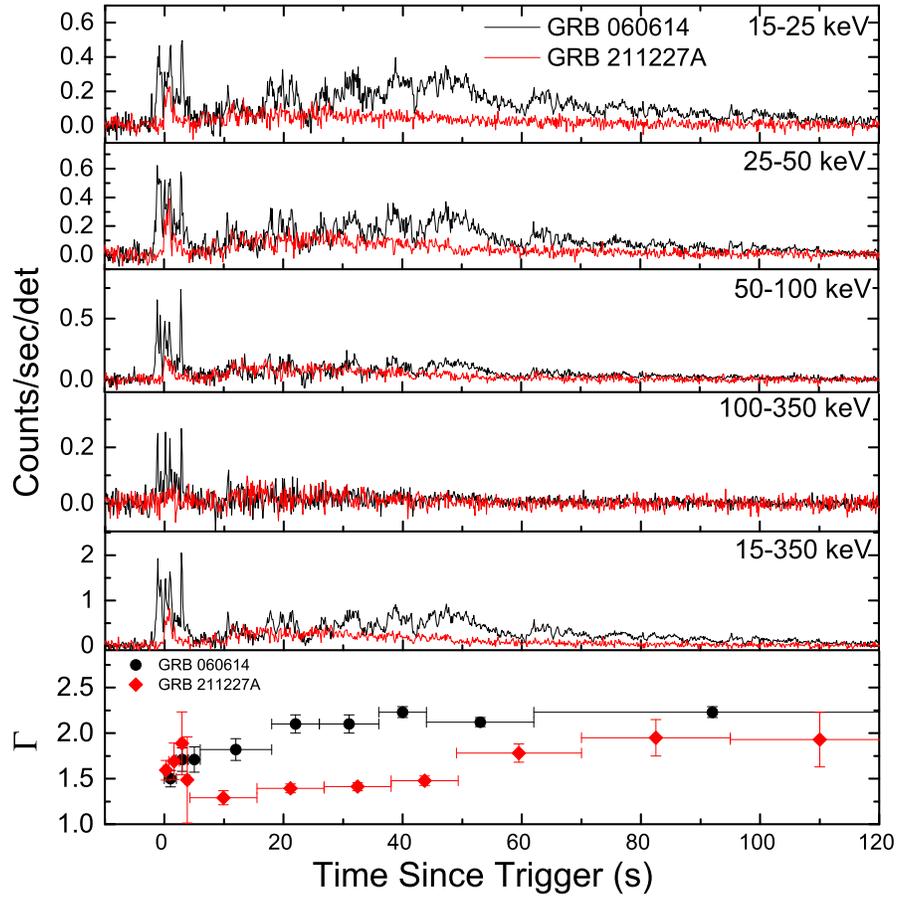}
\caption{{\em Swift}/BAT light curves of GRB 211227A in different
energy bands with a 128 ms time-bin (red solid lines) and the spectral evolution (red diamonds).
In order to compare with GRB 060614, we also plot the light curve (black solid lines) and
spectral evolution (black points) in the same panels.}
\label{fig:BAT}
\end{figure}

GRB 211227A triggered the Burst Alert Telescope (BAT) at 23:32:06 UT on 27 December 2021 (Beardmore
et al. 2021). We downloaded the BAT data from the {\em Swift} website\footnote{$\rm
https://www.swift.ac.uk/archive/selectseq.php?tid=01091101\&source=obs$}, and used the standard
HEASOFT tools (version 6.28) to process the BAT data. For more details on the analysis, please
refer to Sakamoto et al. (2008); Zhang et al. (2009); and L\"{u} et al. (2020). We extract the
light curves in different energy bands with a 128 ms time-bin (Figure \ref{fig:BAT}). The light
curve shows a complex structure with a total duration of about $T_{90}=$84 s, an initial short/hard
spike (with a duration $\sim 4$ s) followed by a series of soft gamma-ray extended emission with a
duration $\sim 80$ s.

\begin{figure}
\centering
\includegraphics [angle=0,scale=0.35] {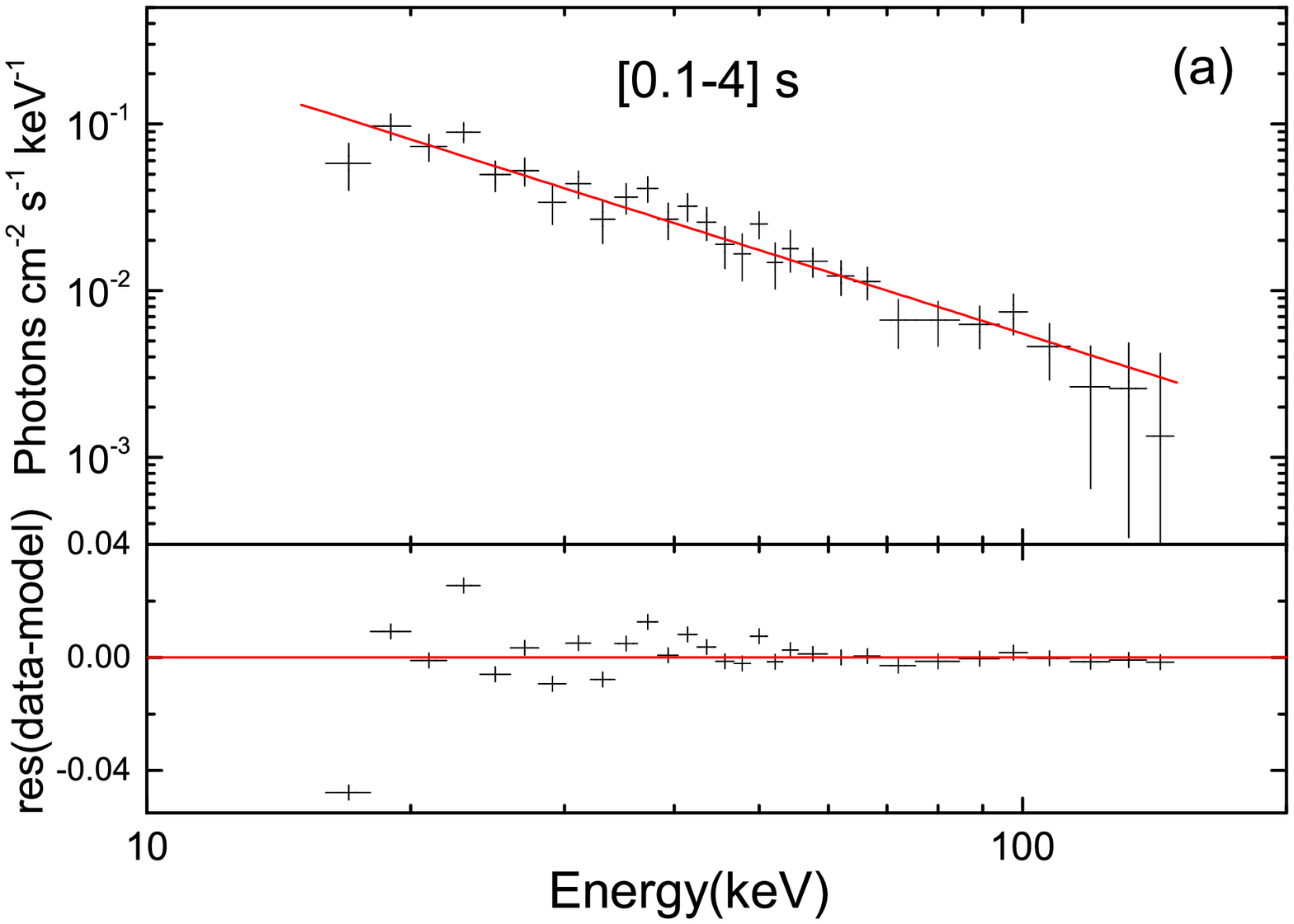}
\includegraphics [angle=0,scale=0.35] {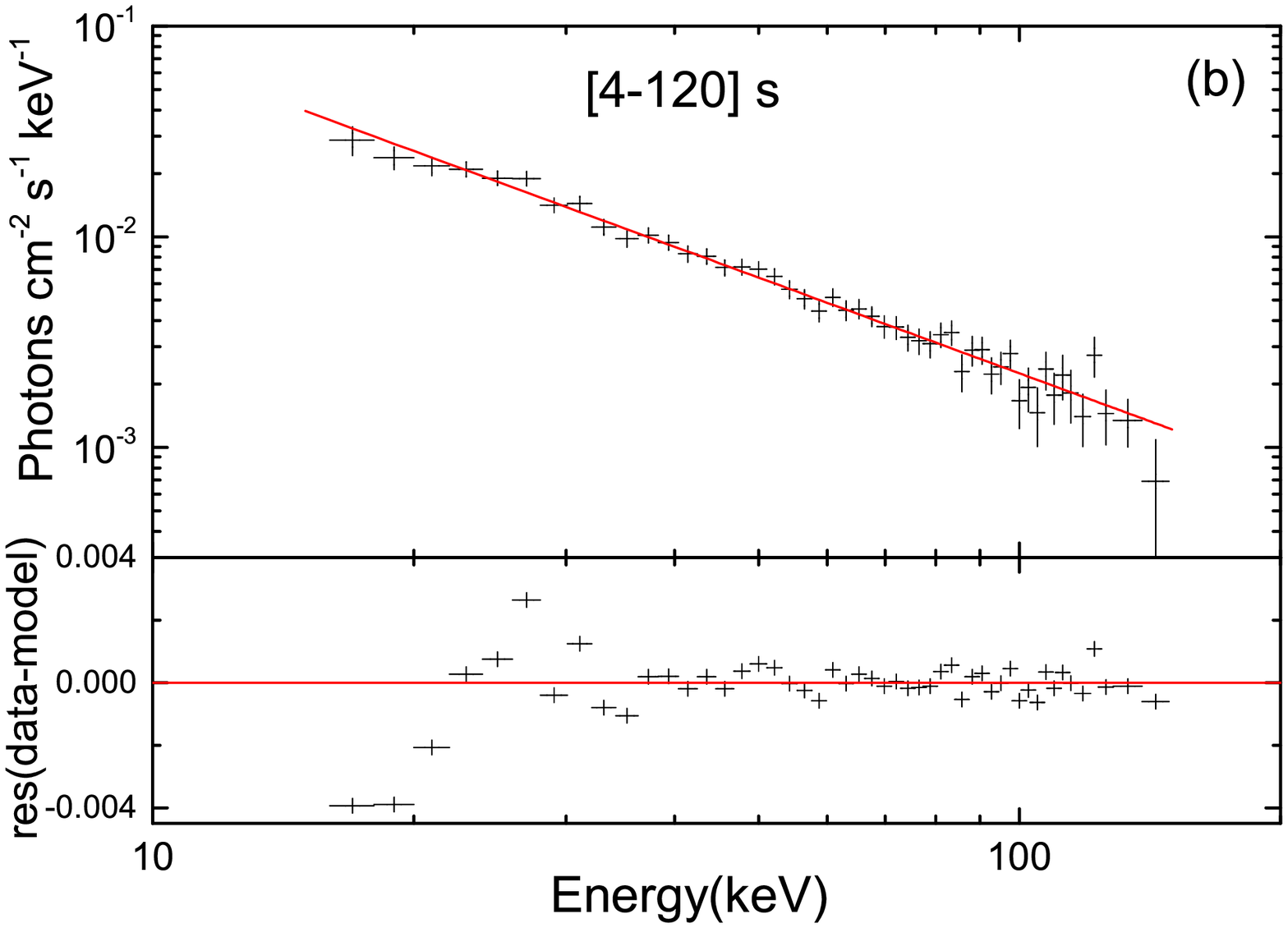}
\includegraphics [angle=0,scale=0.35] {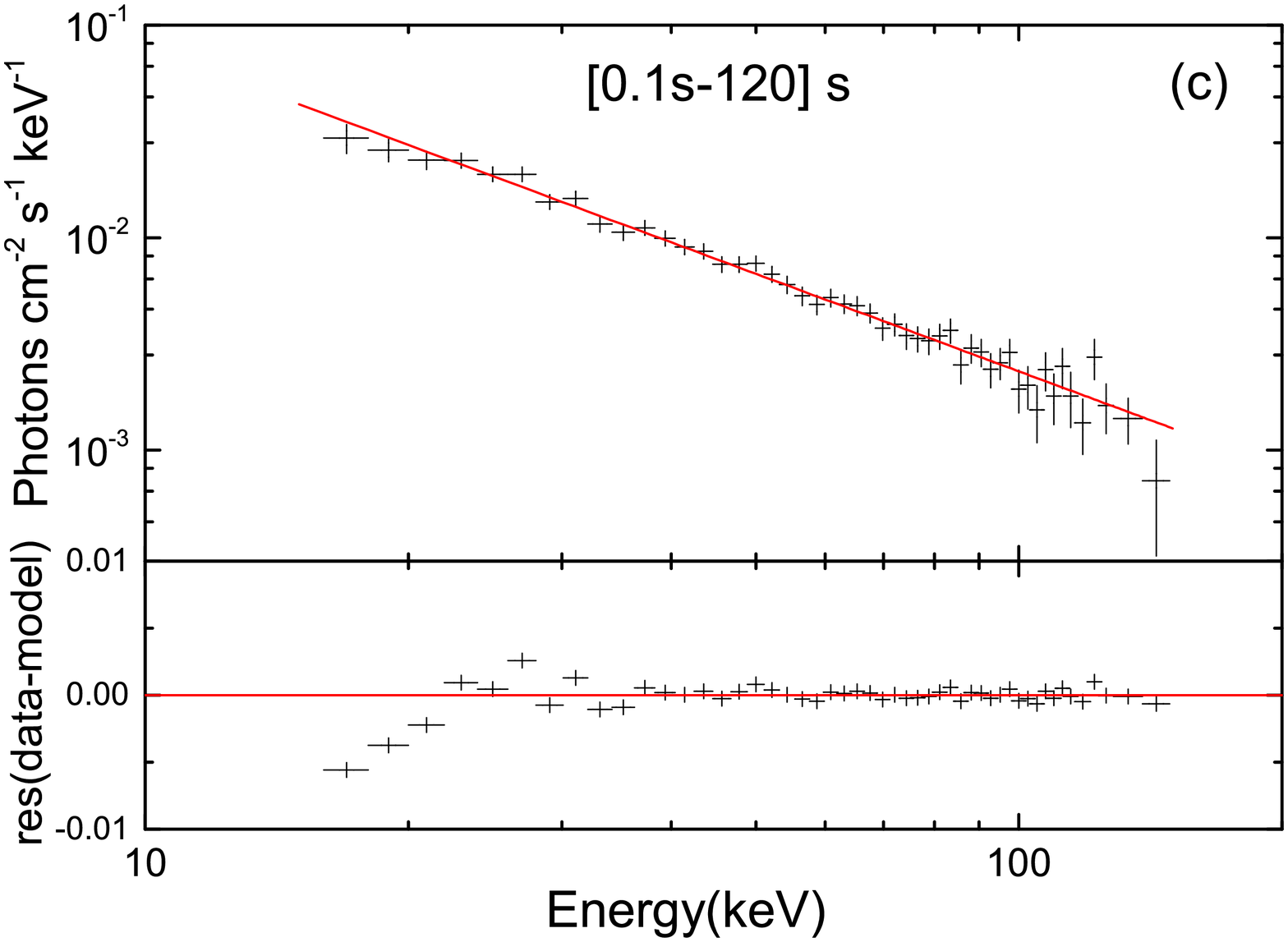}
\caption{The spectral fits of GRB 211227A with power-law model for time-resolved
(a and b) and time-integrated (c).}
\label{fig:spectral}
\end{figure}

We also extract the spectrum by invoking Xspec for fitting, and the background is extracted by
choosing two time-intervals with 80 s before and after the burst. Due to the narrow energy-band of
Swift/BAT, the time-integrated spectrum of the GRB 211227A prompt emission is well fitted by a
simple power-law model ($N\propto E^{-\Gamma}$) with an index $\Gamma=1.53\pm 0.03$. Moreover, we
also separate the light curve into two time slices: the initial hard spike (0.1s-4s) and the
long-lasting extended emission (4s-84s). The spectra of both time slices can be fitted by a
power-law model with $\Gamma_1=1.67\pm 0.08$ and $\Gamma_2=1.5\pm 0.03$, respectively (see Figure
\ref{fig:spectral}). The bottom panel of Figure \ref{fig:BAT} shows a strong temporal evolution of
the spectrum, with $\Gamma\sim1.5$ at the beginning and $\Gamma\sim2$ near the end. The fluence in
the 15-150 keV band is $8.2\pm 0.2\times 10^{-6}~\rm erg ~cm^{-2}$, which corresponds to isotropic
energy $E_{\rm iso}\sim 1.14\times 10^{51}~\rm ergs$ by adopting $z=0.228$ in this energy band.

\subsection{{\em Swift} XRT observations}
\begin{figure}
\centering
\includegraphics [angle=0,scale=0.35] {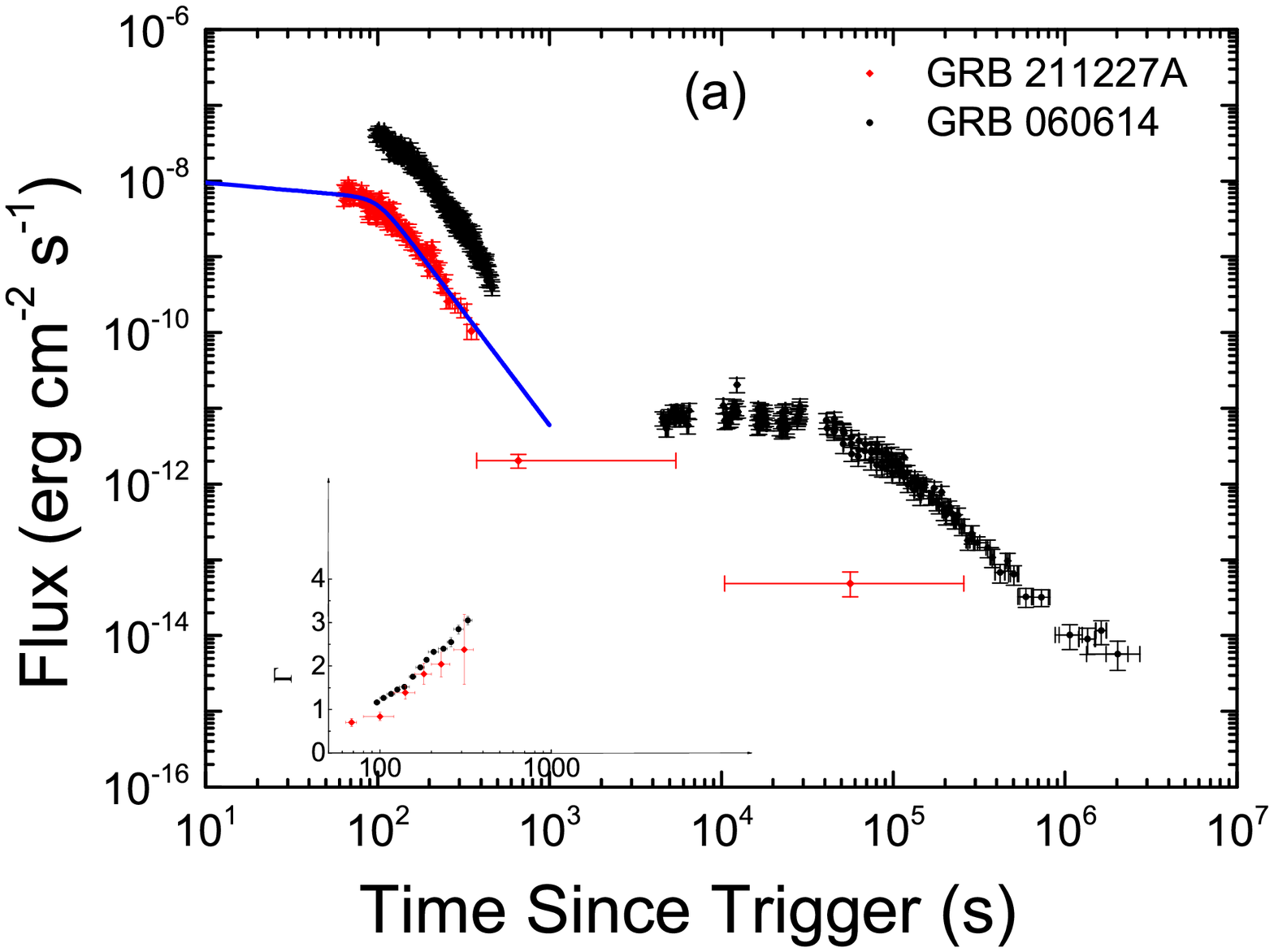}
\includegraphics [angle=0,scale=0.35] {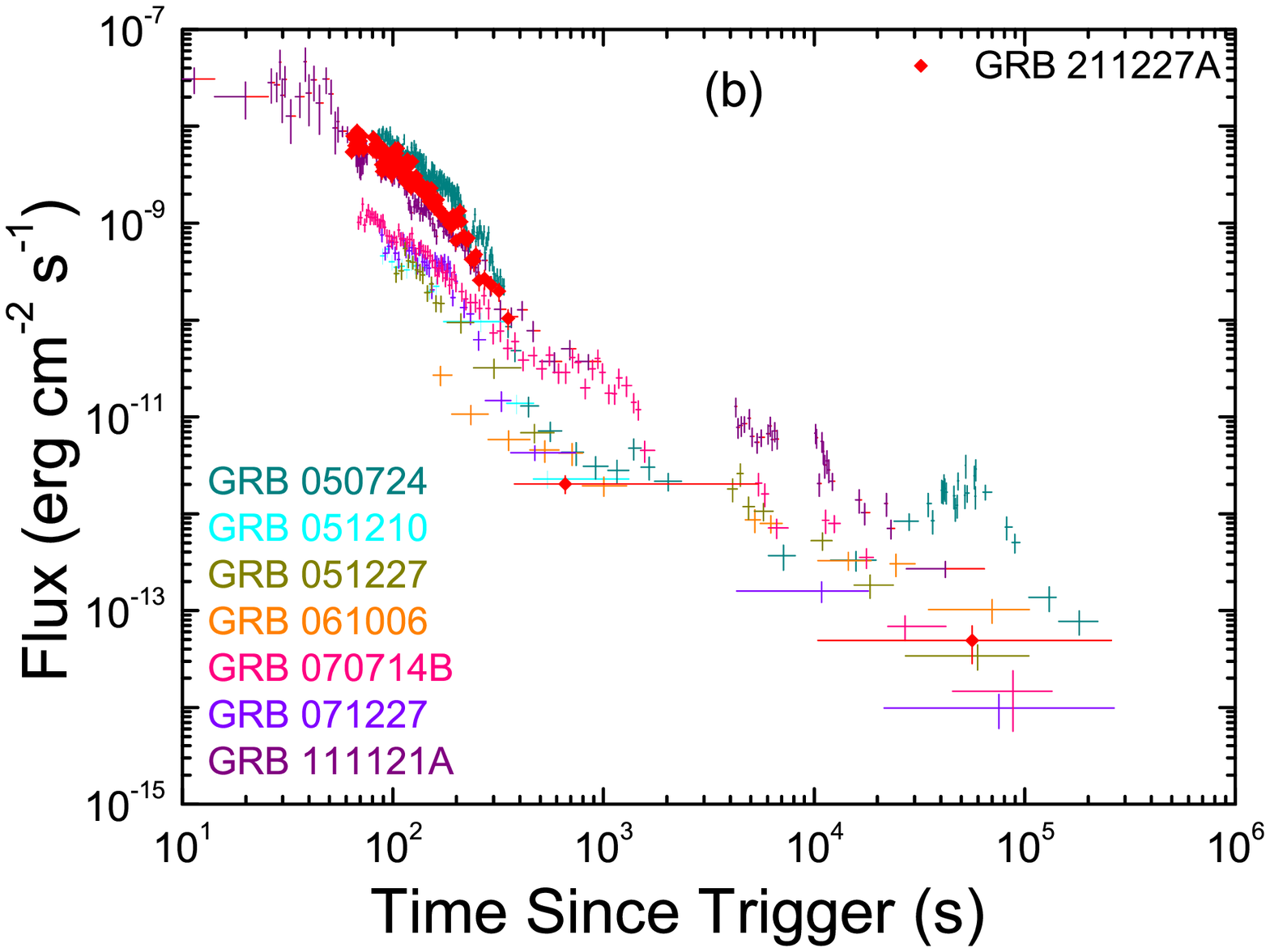}
\caption{X-ray light curve of GRB 211227A (red diamonds). (a): Comparison with GRB 060614
(blue points), broken power law model fits (black solid line), and the spectral evolution (insert). (b):
Comparison with other short GRBs with extended emission and X-ray plateau observed by Swift from L\"{u} et
al. (2015).}
\label{fig:XRT}
\end{figure}

The Swift X-ray telescope (XRT) began observing the field at 73.5 seconds after the BAT trigger. We
made use of the public data from the {\em Swift} archive \footnote{$\rm
https://www.swift.ac.uk/xrt\_curves/01091101$}(Perri et al. 2021). The X-ray light curve in the
early time seems to be a broken power-law decay with indices $\alpha_1=0.2\pm 0.08$,
$\alpha_2=3.10\pm 0.31$, and break time $t_{\rm b}=103\pm12$ s (see Figure \ref{fig:XRT}). We also
extract the time-resolved spectra of the initial X-ray tail with a power law model, and the
spectrum shows a strong temporal evolution that tracked hard-to-soft. The time-integrated spectrum
of the GRB 211227A In X-ray emission is fitted by a simple power-law model with an index
$\Gamma=1.11\pm 0.07$, and the column density of hydrogen $N_{H}$ is $(2.2\pm 0.5)\times
10^{21}~\rm cm^{-2}$.

\subsection{BOOTES follow-up observations}
The Burst Observer and Optical Transient Exploring System (BOOTES) followed GRB 211227A with two
60cm robotic telescopes at BOOTES-2/TELMA station in La Mayora (Malaga, Spain) and BOOTES-4/MET
station at the Lijiang Astronomical Observatory (Yunnan, China). The BOOTES-2/TELMA telescope
performed two epoch observations at 2021-12-27.99UT and 2021-12-28.93UT and the BOOTES-4/MET
telescope took four epoch observations at 2021-12-28.69UT, 2021-12-29.61UT, 2022-01-03.75UT, and
2022-01-04.69UT, respectively. A series of images were obtained using clear and Sloan-i filters
with exposure of 10s, 60s and 90s (Hu et al. 2021). The burst's optical afterglow is not detected
in the stacked images of each epoch which are calibrated via nearby comparison stars from the SDSS
catalogue for the i-filter data or through the transformation equation for the clear filter data.
The obtained upper limits with in 3$\sigma$ are listed in Table\ref{tab:opticaldata}, and the value
of extinction (Av) is 0.062.

\subsection{Lijiang 2.4m Optical Observations}
In order to observe the potential optical emission, on 3 January 2022 ($\sim$6.8 days after
trigger), we observed the field of GRB 211227A with the Lijiang 2.4m telescope, which is located at
the Lijiang Observatory of Yunnan Observatories, Chinese Academy of Science (Wang et al. 2019; Xin
et al. 2020). The photometric observations were performed using the Johnson $R$-band filter with a
total exposure of 1200 s at an airmass of 1.2 and a seeing of $\sim$1.7 under good weather
conditions. Three nearby faint stars are in the field of view (FOV) which can provide calibration
of the target. However, we did not detect any source to a depth of R$>$21.5 mag within the XRT
enhanced error box. On 7 January 2022 ($\sim$10.7 days after trigger), we observed this field
again, with a total exposure time to 1800 s. Again, no signal was detected, and only an upper limit
with R$>$21.8 mag was obtained.

\subsection{Other telescopes follow-up observations}
Beside the BOOTES and Lijiang 2.4m, there were several ground-optical telescopes that performed
follow-up observations of the source after the GRB 211227A trigger, such as the Nanshan/NEXT-0.6m
located at Nanshan, Xinjiang, China (Fu et al. 2021), the LCO 1-m Sinistro instrument at the Teide,
Tenerife site (Strausbaugh \& Cucchiara 2021), the 2m robotic Liverpool telescope (Perley 2021), the CAHA
2.2m telescope at the Calar Alto Observatory, Almeria, Spain (Kann et al. 2021), and the
Gemini/GMOS-S (O'Connor et al. 2022). However, none of them could detect any source within upper
limits.

\subsection{Host galaxy Observation}
\begin{figure}
\centering
\includegraphics [angle=0,scale=0.25] {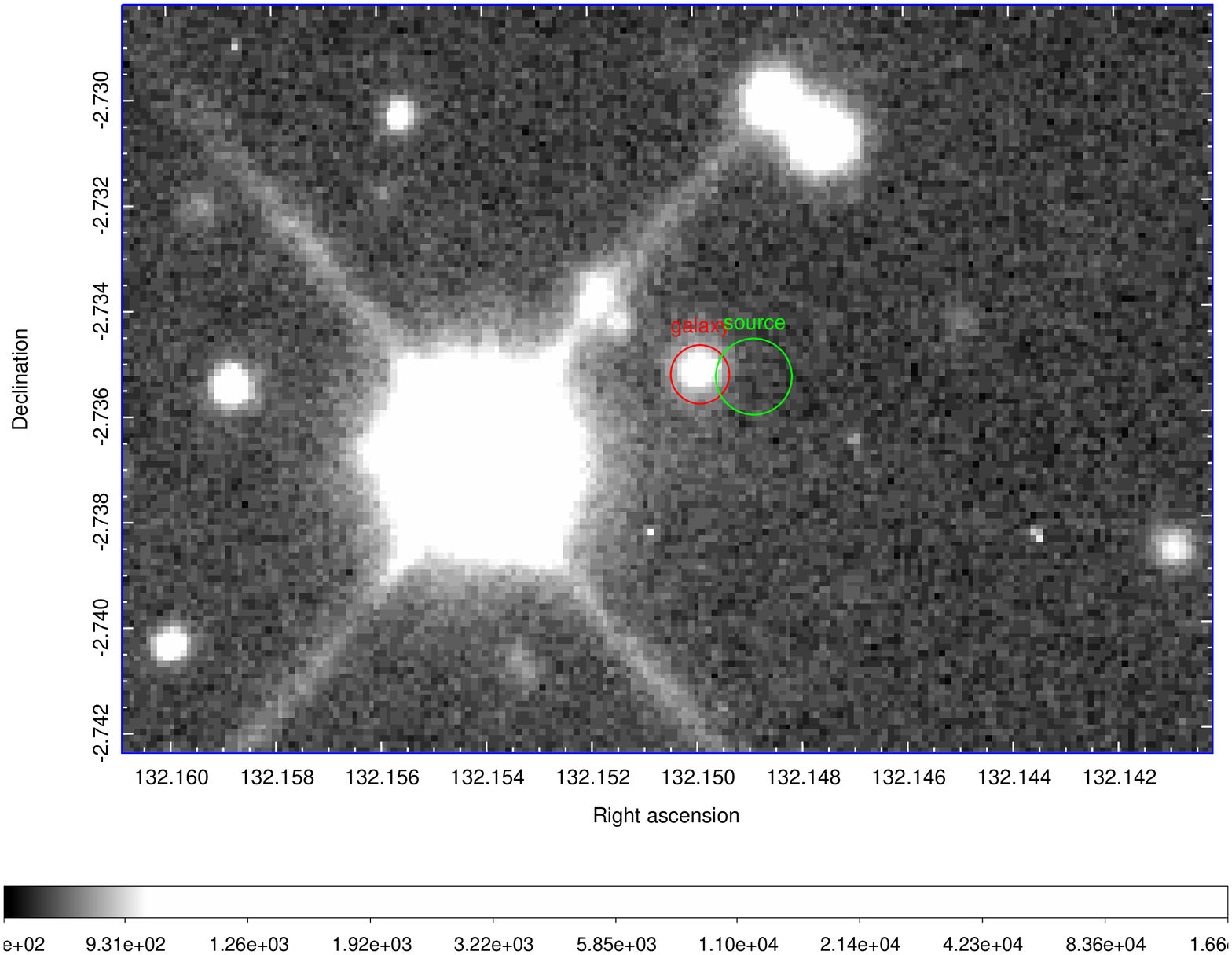}
\includegraphics [angle=0,scale=0.37] {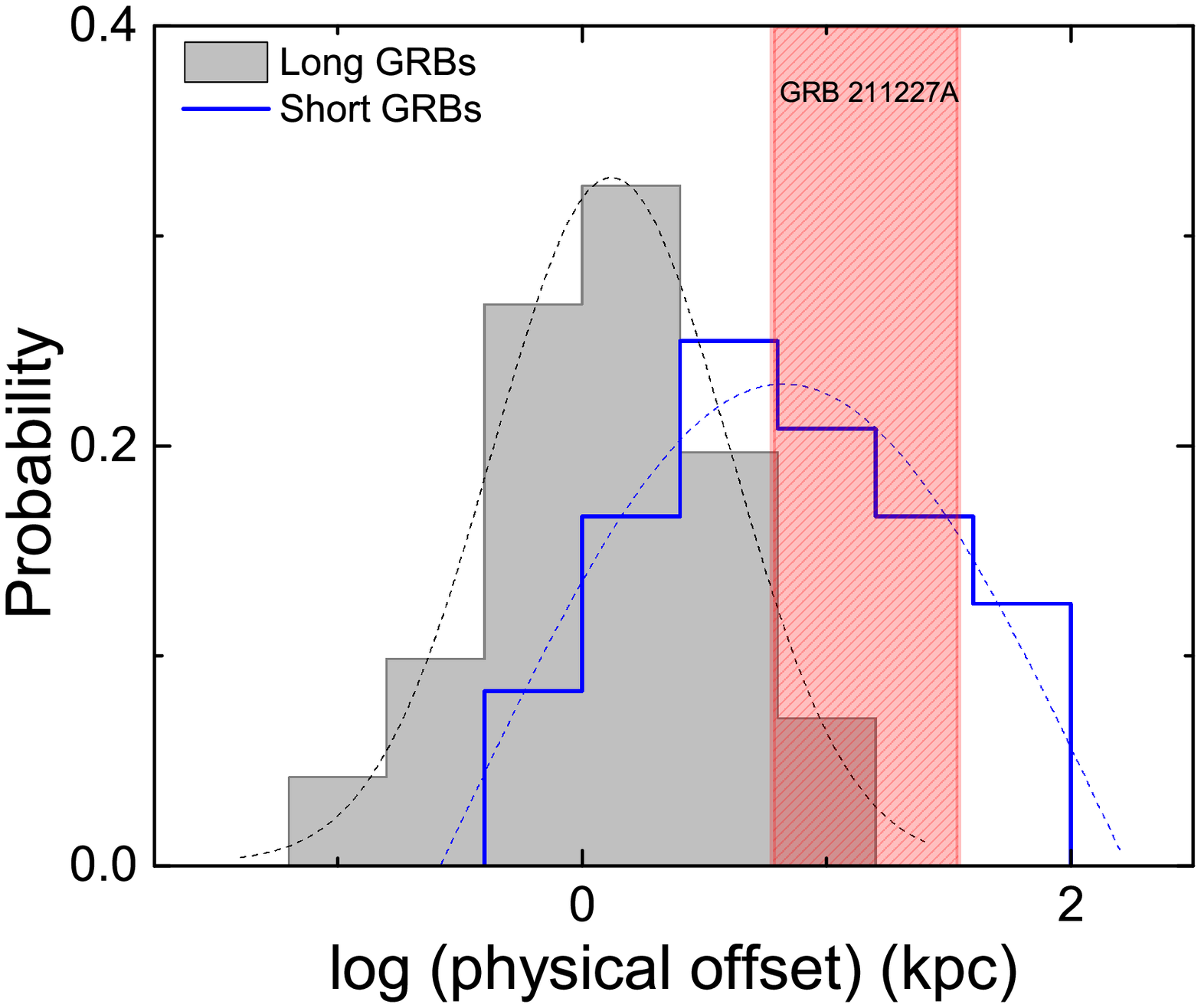}
\caption{Left: The follow-up observation of GRB 211227A with LCOGT,
and the red circle is the location (RA$=$08:48:35.975, Dec$=$-02:44:06.93)
of the galaxy with a radius of 2 arcseconds, and the green circle is the
location (RA$=$08:48:35.73, Dec$=$-02:44:07.1) of the GRB 211227A with an
error-radius of 2.6 arcseconds. Right: Distributions of the physical offsets
of long and short GRBs taken from L\"{u} et al. (2017), as well as GRB 211227A (vertical solid red line).
The dashed lines are the best Gaussian fits for long and short GRBs, respectively.}
\label{fig:Host}
\end{figure}

We observed the localization of GRB 211227A by using the 1m telescope at the Las Cumbres
Observatory Global Telescope (LCOGT) at the Teide Tenerife site. The observation started at
01:28:37 UT on 28 December 2021 (from 6991s to 8599s after the BAT trigger), and consisted of
5$\times$300 s integrations in the $R$ band. Our combined image is centered at 7795s after the
trigger. No afterglow is detected in the XRT error circle in our stacked image down to a limiting
magnitude of R$>23.81$ mag, calculated with respect to the USNO-B.1 catalog. However, there exists
a known source at the border of the Swift/XRT error circle at coordinates: R.A.
(J2000)=08:48:35.975 and Dec (J2000)= -02:44:06.93, which has R $\sim 19.13\pm0.09$ mag from our
images, classified as a galaxy with a photo-redshift of $z=0.244\pm0.089$ from SDSS (Figure
\ref{fig:Host}). Moreover, Malesani et al. (2021) reported that 5 emission lines are observed in
the putative host galaxy (e.g., [O II], Hbeta, [O III], Halpha, [N II], and [S II]), and they
proposed that a possible host galaxy of GRB 211227A with redshift $z$=0.228 was detected by the
X-shooter spectrograph. The redshift $z=0.244\pm0.089$ what we measured is consistent with that
reported in Malesani et al. (2021) with $z=$0.228 via the X-shooter spectrograph, and the position
of this source for LCOGT observations is the same with GCN report in Malesani et al. (2021).  In
our calculations, we adopt $z=$0.228 which is the spectroscopic redshift. In addition, we also
calculate the physical distance from the center of GRB 211227A to the host galaxy center as
$20.47\pm14.47$ kpc if we assume that the galaxy of $z=$0.228 is intrinsic host of GRB 211227A.


\section{Comparison with GRB 060614 and other short GRBs}
From the observational point of view, the duration of GRB 211227A is much longer than that of
short-duration GRBs, but is consistent with that of typical long-duration GRBs. Even only
considering its initial hard spike lasting $\sim$4 s, is still characterized this burst as a
category of long-duration GRBs. However, based on the characteristics of the GRB 211227A prompt
emission, it is very similar to that of GRB 060614 which had an initial short/hard spike with
$\sim$5 s followed by a series of soft gamma-ray extended emission lasting $\sim$102 s (Gal-Yam et
al. 2006; Zhang et al. 2007a). Figure \ref{fig:BAT} shows the complete comparison of the prompt
emission light curves in different energy bands. The profile of the light curves of GRB 211227A and
GRB 060614 are consistent with each other, and both of them are characterized by a short/hard spike
followed by a series of soft gamma-ray extended emission with a duration $\sim 100$ s. Moreover, we
also present their spectral evolution, and find that a strong temporal evolution is obvious in
both. The evolution behavior of the spectrum is from $\Gamma\sim1.5$ at the beginning to
$\Gamma\sim2$ near the end. Moreover, the isotropic energy of GRB 211227A is $E_{\rm iso}\sim
1.14\times 10^{51}~\rm ergs$, which is also close to that of GRB 060614 with $E_{\rm iso}\sim
7.76\times 10^{50}~\rm ergs$.

In terms of the X-ray afterglow, the X-ray emission of GRB 211227A is not significant difference
with other long-duration GRBs that presented several segments (Zhang et al. 2006). The spectrum
shows a strong temporal evolution that tracked hard-to-soft, which is consistent with the behavior
of X-ray tail emission in most long- and short-duration GRBs (Zhang et al. 2007b). Figure
\ref{fig:XRT} shows the X-ray emission of GRB 211227A in comparison to that of GRB 060614 (left)
and other short GRBs with extended emission (right). We find that the temporal decay index of the
X-ray tail before $10^3$ s is consistent with that of GRB 060614 ($f\propto t^{-3}$; Zhang et al.
2007b), and shows a similar temporal evolution that tracked from hard-to-soft. Moreover, we also
collect all of the X-ray light curves of the short GRBs with extended emission and plateau emission
in X-ray observed by Swift (L\"{u} et al. 2015)\footnote{There are 7 short GRBs with both extended
emission and X-ray plateau emission (e.g., GRBs 050724, 051210, 051227, 061006, 070714B, 071227,
and 111121A), and the fraction is about $6\%$ of total short GRBs.}, and compare with the X-ray
emission of GRB 211227A. We find that the behavior of the early X-ray emission of GRB 211227A is
also similar to that of other short GRBs with extended emission.

Moreover, Figure \ref{fig:Host} shows the comparison of the physical offset of GRB 211227A with
that of other long and short GRBs. We find that the physical offset is larger than that of most
long-duration GRBs, but is consistent with the distribution of other short GRBs within a larger
error range.

\section{Physical origin: massive star core-collapse or compact star merger?}
In general, long-duration GRBs are believed to originate from the deaths of massive stars, and
their host galaxies are typically irregular galaxies with intense star formation (Woosley 1993;
Fruchter et al. 2006). In contrast, short-duration GRBs are associated with nearby early-type
galaxies with little star formation and are related to compact star mergers with a large offset
from the center of the host galaxy (Berger et al. 2005; Fong et al. 2010). Moreover, from the
statistical point of view, Leibler \& Berger (2010) found that the two GRB host populations remain
distinct based on the stellar mass distribution and population ages of galaxies. However, such a
cozy picture was destroyed by nearby long-duration GRB 060614 without a SN association (Gehrels et
al, 2006; Gal-Yam et al, 2006; Fynbo et al, 2006; Della Valle et al, 2006) and short-duration GRB
200826A associated with a SN (Zhang et al. 2021; Ahumada et al. 2021). Interestingly, GRB 211227A
shares some similar properties with GRB 060614, and has a large physical offset from the galaxy
center. One question is what is the physical origin of GRB 211227A? In this section, we attempt to
find some clues to reveal the physical origin of this case.

\subsection{Massive star core-collapse scenario}
\begin{figure}
\centering
\includegraphics [angle=0,scale=0.6] {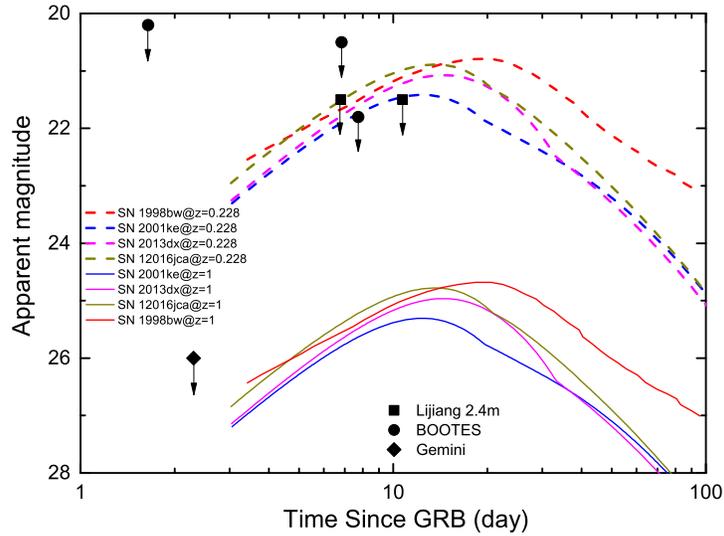}
\caption{Comparison of the observed data of the GRB 211227A from the Lijiang 2.4m and BOOTES
with those of SN 1998bw, SN 2001ke, SN 2013dx, and SN 2016jca associated with
GRBs at $z=$0.228 (dashed lines) and $z=$1 (solid lines).}
\label{fig:SN}
\end{figure}

The ``smoking gun" signature of a long-duration GRB from a massive star core-collapse is the
detection of the associated SN in the optical band (Galama et al. 1998; Kippen et al. 1998; Hjorth
et al. 2003; Stanek et al. 2003; Malesani et al. 2004; Della Valle et al. 2006; Mazzali et al.
2008; Bufano et al. 2012; Xu et al. 2013; Cano et al. 2017; L{\"u} et al. 2018). Being a
long-duration GRB 211227A at a low redshift $z=$0.228, it is expected that a SN component should be
detected around ten days after the trigger, yet it is surprising that deep searches of an
underlying SN give null results. This raises interesting questions regarding whether the distance
is too large for detecting a less energetic event. Therefore, we shift several SNe to the redshift
$z=$0.228 to see how bright they are. Figure \ref{fig:SN} shows light curves of SN 1998bw, SN
2001ke, SN 2013dx, and SN 2016jca associated with some long-duration GRBs at $z=$0.228. We find
that the upper limit of the luminosity of the SN (also the limiting magnitude of the Lijiang 2.4m
telescope with the limit luminosity as $(4.3-5.6)\times 10^{42}~\rm erg ~s^{-1}$), if any, is
several times fainter than SN 1998bw and fainter than other Type Ic SN associated with GRBs at
$z=$0.228 (Galama et al. 1998; Della Valle et al. 2003; Xu et al. 2013; Cano et al. 2017). It
suggests that lack of a SN associated with GRB 211227A is not caused by too large of a distance,
but more likely physical reasons. In order to compare those SNe at $z=$0.228 with that of higher
redshift, we re-plot those SNe if we put them into $z=$1. It is much dimmer than the upper limits
of observations. On the other hand, based on the upper limit of observations, one can roughly
estimate the redshift $z=0.285$ of SNe that can be ruled out. Moreover, the measurement of a large
physical offset is also inconsistent with that of other long GRBs. From the prompt emission
analysis, on the other hand, GRB 211227A is consistent with the properties of GRB 060614. Those
facts suggest that it does not likely match with the physical properties expected in massive star
core-collapse. If so, the lack of a SN signature is a natural expectation. However, the origin of
massive star collapse of GRB 211227A is still to be ruled out if it occur in the host galaxy with a
higher redshift.

\subsection{Compact star mergers scenario}
\begin{figure}
\centering
\includegraphics [angle=0,scale=0.6] {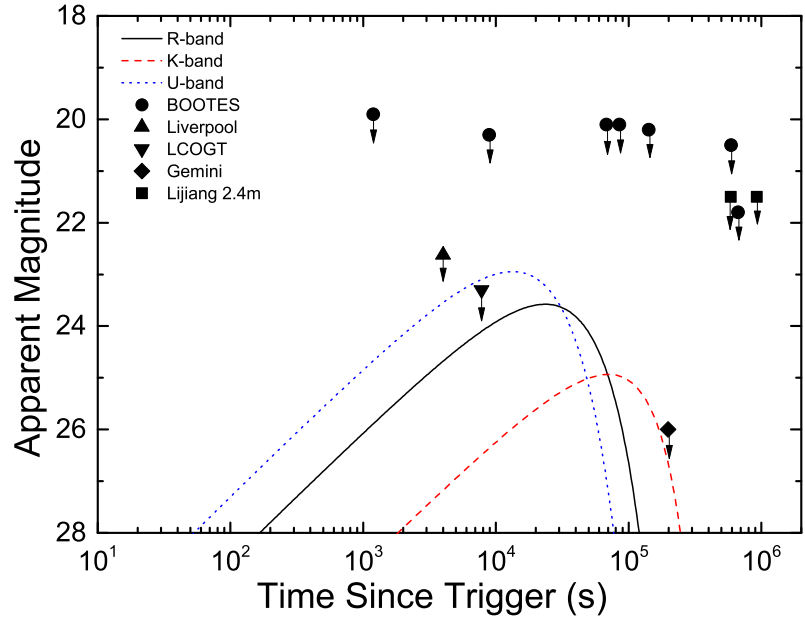}
\caption{Calculated kilonova emission of GRB 211227A at $z=$0.228 by considering one energy source
with $M_{ejc}\sim0.03 M_{\odot}$, $\beta\sim 0.25c$, and $\kappa\sim 0.97~\rm
cm^{2}~g^{-1}$ which are taken from kilonova AT2017gfo (Yu et al. 2018).}
\label{fig:KNNS}
\end{figure}

The direct detection of GW170817 and its electromagnetic counterpart (e.g., GRB 170817A and
AT2017gfo), which was the first identified multimessenger gravitational-wave and electromagnetic
signal, originated in the merger of a binary NS system (Abbott et al. 2017; Covino et al. 2017;
Goldstein et al. 2017; Savchenko et al. 2017; Zhang et al. 2018). The observations of kilonova
AT2017gfo were comprehensive, long lasting, and multi-wavelength compared to previous kilonova
candidates (Coulter et al. 2017; Arcavi et al. 2017; Lipunov et al. 2017; Tanvir et al. 2017;
Valenti et al. 2017; Ai et al. 2018; Metzger 2019; Rossi et al. 2020). A roughly constraint of
kilonova parameters can be realized via a multi-wavelength fit with the kilonova model (Yu et al.
2018; Li et al. 2018; Ai et al. 2018). By adopting the same parameters ($M_{ejc}\sim0.03
M_{\odot}$, $\beta\sim 0.25c$, and $\kappa\sim 0.97~\rm cm^{2}~g^{-1}$) with kilonova AT2017gfo,
one can also calculate the kilonova emission of GRB 211227A in K-, R-, and U-bands at redshift
$z=$0.228 (see Figure \ref{fig:KNNS}). The peak magnitude is not much of an improvement, and it
remains fainter than the observed limitation.

\begin{figure}
\centering
\includegraphics [angle=0,scale=0.6] {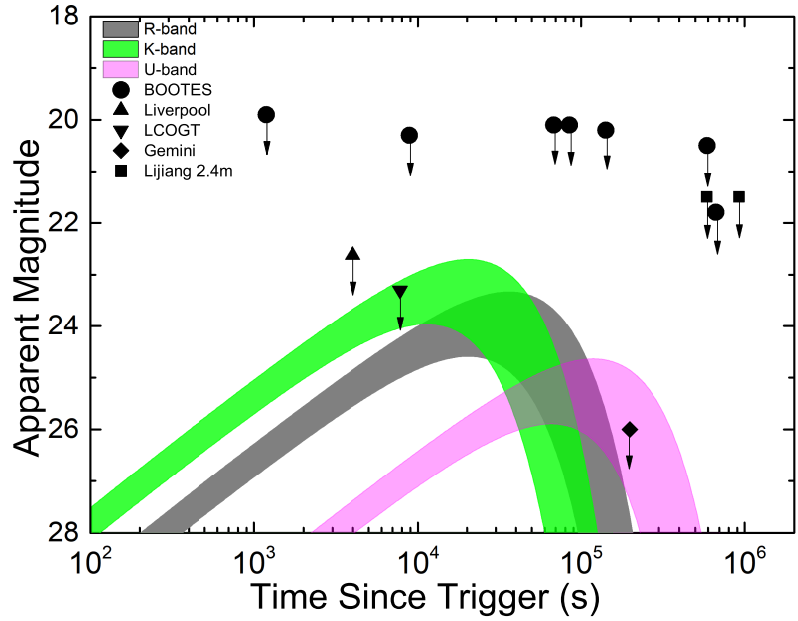}
\caption{Light curve of kilonova in $K$, $r$, and $U$ bands of GRB 211227A by adopting
$M_{ejc}\sim(0.01-0.1) M_{\odot}$, $\beta\sim 0.2c$, and $\kappa\sim 10~\rm cm^{2}~g^{-1}$ at $z=$0.228.}
\label{fig:KN}
\end{figure}

Yang et al. (2015) reported the discovery of near-infrared bump that is inconsistent with the
afterglow emission of GRB 060614, but arises from a kilonova. They invoke the kilonova model to fit
the excess in the near-infrared band with the ejecta mass ($M_{ejc}\sim0.1 M_{\odot}$), velocity
($\beta\sim 0.2c$), and opacity ($\kappa\sim 10~\rm cm^{2}~g^{-1}$). {In order to compare the
kilonova emission between GRB 211227A and GRB 060614,} we calculate the possible kilonova emission
by adopting the same parameters with that of GRB 060614 even with the large uncertainty. Figure
\ref{fig:KN} shows the kilonova emission in K-, R-, and U-bands at redshift $z=$0.228 by
considering only one energy source. Here, the ejecta mass range $(0.01-0.1) M_{\odot}$) is adopted
in our calculations. The peak magnitude of kilonova emission is much lower than the observed
limitation of the Lijiang 2.4m and current other optical telescopes. If so, the lack of a kilonova
signature is the natural expectation for a less energetic event.

Inspired by GRB 170817A and GRB 060614, we propose that the long-duration GRB 211227A is from a
more energetic merger event. Moreover, the large physical offset ($\sim$20.47 kpc) of this case
that is larger than that of most long GRBs, is also supported this hypothesis even still consistent
with serval long GRBs within a large error. In general, robust associations of a
fainter-than-supernova optical/IR transient (called kilonova) with some GRBs suggest that they are
likely related to NS-NS or NS-BH mergers (Tanvir et al. 2013; Berger 2014; Yang et al. 2015; Abbott
et al. 2017; Goldstein et al. 2017; Savchenko et al. 2017; Troja et al. 2017; Zhang et al. 2018).
Such a transient is powered by the r-process (rapid neutron capture) and radioactive decay of the
synthesized heavy elements (e.g., Li \& Pacz\'{y}nski 1998; Freiburghaus et al. 1999; Hotokezaka et
al. 2013; Coulter et al. 2017; Arcavi et al. 2017; Lipunov et al. 2017; Tanvir et al. 2017; Valenti
et al. 2017). If GRB 211227A is from a compact star merger, one can roughly calculate how bright it
is by adopting some typical parameters.

For a NS-NS merger, additional energy should be injected into the ejecta if the post-merger remnant
is a NS (Metzger et al. 2010; Yu et al. 2013; Gao et al. 2017; Yu et al. 2018; Ai et al. 2018; Yuan
et al. 2021). We also invoke a model of hybrid energy sources (r-process and energy injection from
the NS) to calculate the kilonova emission. L\"{u} et al. (2015) proposed that both extended
emission and internal plateau can be produced by magnetar spin-down\footnote{In other words, the
extended emission is essentially the brightest internal plateau commonly observed in short GRBs,
and a more sensitive and softer detector would detect more EE from short GRBs}. The observed EE and
X-ray tail of GRB 211227A seems to come from the contribution of a magnetar spin-down. If this is
the case, we adopt the same parameters as above in the r-process, and $L_0=10^{48}\rm erg~s^{-1}$
and $t=105$ s as the initial luminosity and time scale of energy injection from NS. We find that
the peak magnitude is also not improved even when considering hybrid energy sources, because the
time-scale of energy injection is too short to be contributed to the kilonova. Hence, the
progenitor of GRB 211227A seems to come from compact star mergers, which can naturally explain the
lack of SN and kilonova emission, the characteristic of prompt emission, the initial X-ray
emission, and the large physical offset.

\section{Conclusion and discussion}
GRB 211227A was observed by Swift to have a duration of $\sim 84$ seconds at redshift $z=$0.228,
but the light curve is characterized by an initial short/hard spike (with a duration $\sim 4$ s)
followed by a series of soft gamma-ray extended emission with a duration $\sim 80$ s. Both the
prompt emission and the X-ray initial tail of this case show a strong temporal evolution that
tracked from hard-to-soft, and those behaviours are very similar to that of GRB 060614. Several
optical telescopes made follow-up observations of the afterglow and host galaxy, but did not detect
any afterglow emission except for upper limitations. We also applied for the Lijiang 2.4m optical
telescope to observe this event twice, but obtained only upper limits. Based on the LCOGT
observations, we can roughly estimate the physical offset from the galaxy center as $20.47\pm14.47$
kpc.

At such low redshift, it is expected that a SN component should be detected around ten days after
the trigger if we believe the massive star core-collapse origin, yet it is surprising that deep
searches of an underlying SN give null results. We shift several SN events that are associated with
long-duration GRBs into the redshift $z=$0.228, and find that the upper limit of the luminosity of
the SN, if any, is several times fainter than SN 1998bw, and fainter than other Type Ic SN
associated with GRBs at $z=$0.228. The lack of an associated SN and the large physical offset of
this case suggest that the progenitors of GRB 211227A do not seem to originate from a massive star
core-collapse. If so, the lack of a SN signature and the large physical offset are a natural
expectation.

Alternatively, we propose that the long-duration GRB 211227A is from a compact star merger. If this
is the case, we also calculate the kilonova emission by adopting the same parameters with that of
GRB 060614 and GRB 170817A, and find that the kilonova emission is too faint to be detected. Hence,
the progenitor of GRB 211227A seems to come from compact star mergers, which can naturally
interpret the lack of SN and kilonova emissions, the characteristics of the prompt emission, and
the large physical offset. Moreover, Kann et al. (2011) used a large sample of GRB afterglow data
to compare the optical afterglows of Type I (originated from compact star merger) and Type II GRBs
(related to death of massive star), and found that the optical afterglow of Type I GRBs are
intrinsically lower than that of Type II GRBs. The lack of optical afterglow of GRB 211227A also
support its origin in a compact star merger.

On the other hand, GRB211227A is a possible optically-dark burst. Melandri et al. (2012) studied
the properties of the population of completely optically-dark GRBs that observed by {\em Swift},
and presented the relationship between optical flux and X-ray flux that observed at $t=11$ hours
for completely sample of optically-dark GRBs. In order to test this possibility, we also calculate
the upper limits of both optical and X-ray flux observed at $t=11$ hours, and find that the
observed both optical and X-ray flux at $t=11$ hours are much lower than that of optically-duck
bursts identified in Melandri et al. (2012). So that, the possibility that GRB 211227A is an
optically-dark GRB cannot be ruled out. If this is the case, GRB 211227A should be generated in
much denser environments in the host galaxy. Moreover, another possibility is that the GRB 211227A
is not host at $z$=0.228, but located at a higher redshift with a faint and underlying galaxy which
we do not observe by the deepest upper limit of the field. Bloom et al (2002) and Lyman et al.
(2017) pointed out that one can calculate the probability of chance coincidence of possible host
galaxy. Following the method of those two paper, we also roughly estimate the probability of chance
coincidence $P_{\rm ch}=0.038$, which is less than 0.1 ($P_{\rm ch}\geq0.1$ no obvious host; Bloom
et al (2002)). So that, the galaxy at $z$=0.228 seems to be the host of GRB 211227A.

In order to investigate the progenitor of a GRB 211227A-like event, more information on the host
galaxy is essential. We therefore encourage intense multi-band, more sensitive optical follow-up
observations of GRB 211227A-like events to catch electromagnetic signals (e.g., kilonova or SN
signatures) and the properties of the host galaxy in the future. If possible, the observed GRB
211227A-like event with a GW signal will also provide a good probe to study progenitors.

\begin{acknowledgements}
We are very grateful to thank the referee for constructive comments and suggestions to improve this
manuscript. We also thank WeiKang Zheng and Meng-Hua Chen for helpful comments. We acknowledge the
use of the public data from the Swift data archive. This work is supported by the National Natural
Science Foundation of China (grant Nos. 11922301, 12133003, and U1938201), the Guangxi Science
Foundation (grant Nos. 2017GXNSFFA198008, and AD17129006), the Program of Bagui Young Scholars
Program (LHJ), the teaching reform project of Guangxi Higher education(2019JGZ102), and special
funding for Guangxi distinguished professors (Bagui Yingcai and Bagui Xuezhe). AJCT acknowledgdes
support from the Spanish Ministry project PID2020-118491GB-I00 and the collaboration with the rest
of IAA-CSIC ARAE group and the UMA Unidad Asociada to CSIC colleagues. MCG ackowledges support from
the Ram\'on y Cajal Fellowship RYC2019-026465-I. YDH acknowledges support under the additional
funding from the RYC2019-026465-I.

\end{acknowledgements}


\begin{table*}
\centering \caption{Optical observations of GRB211227A with BOOTES, LCOGT, and Lijiang 2.4m in
3$\sigma$. Not corrected for galaxy extinction.}
\begin{tabular}{cccccccc}
			\hline
			$T_{\rm start}$& $T_{\rm end}$ &$T_{\rm mid}(s)$& Telescope & Band  &Exposure & Magnitude\\
			2021-12-27T23:42:44&2021-12-28T00:01:04&1188&BOOTES-2&clear&$40\times10$&$>19.9$\\
			2021-12-28T00:36:39&2021-12-28T03:23:28&8877&BOOTES-2&clear&$64\times60$&$>20.3$ \\
			2021-12-28T22:17:41&2021-12-29T00:03:14&85101&BOOTES-2&clear&$41\times60$&$>20.1$ \\		
			2021-12-28T16:32:19&2021-12-28T20:18:40&68003&BOOTES-4&clear&$92\times60$&$>21.1$ \\
			2021-12-29T14:43:21&2021-12-29T15:16:54&142081&BOOTES-4&clear&$26\times60$&$>20.2$ \\
			2022-01-03T17:56:17&2022-01-03T20:57:59&590102&BOOTES-4&i&$18\times90$&$>20.5$ \\
			2022-01-04T16:30:15&2022-01-04T17:53:52&668397&BOOTES-4&i&$23\times90$&$>21.8$ \\
            \hline
            2021-12-28T01:28:37&2021-12-28T01:55:25&7795&LCOGT&R&$5\times300$&$>23.81$ \\
            \hline
            2022-01-03T18:23:25&2022-01-03T18:43:27&587520&Lijiang-2.4m&R&$1\times1200$&$>21.5$ \\
            2022-01-07T15:14:28&2022-01-07T15:44:45&924480&Lijiang-2.4m&R&$2\times900$&$>21.8$ \\
            \hline
            \multicolumn{6}{c}{}\\[4pt]
	
\end{tabular}
\label{tab:opticaldata}		
\end{table*}

\clearpage



\end{document}